\documentclass[journal]{IEEEtran}

\usepackage{float}
\usepackage{caption}
\usepackage{subcaption}
\usepackage{longtable}
\usepackage{tabularx}
\usepackage{tabu}
\usepackage{multirow}
\usepackage{tikz}
\usepackage{graphicx}
\usepackage{tikzscale}
\usepackage{pgfplots}
\usepackage{setspace}
\usepackage{booktabs}
\usepackage{siunitx}
\usepackage{booktabs}
\usepackage{adjustbox}
\usepackage{makecell}
\usepackage{color, soul}
\usepackage{booktabs}
\usepackage{threeparttable}

\usepackage{algorithmic}
\usepackage{amsmath,amssymb,amsfonts}
\usepackage[ruled,linesnumbered]{algorithm2e}
\usepackage[font=small,labelfont=bf]{caption}
\SetKwFor{Foreach}{for each}{do}{endForEach}%
\SetKwFor{For}{for}{do}{endfor for}%
\SetKwIF{uIf}{ElseIf}{Else}{if}{then}{else if}{else}{endIf if}%
\SetKwFor{While}{while}{do}{endWhile while}%

\begin{document}

\title{
Trustworthy Privacy-preserving Hierarchical Ensemble and Federated Learning in Healthcare 4.0 with Blockchain
}

\author{%
Veronika~Stephanie, Ibrahim~Khalil, Mohammed~Atiquzzaman  \IEEEmembership{Senior Member, IEEE}, and Xun~Yi
}

% \markboth{IEEE TRANSACTIONS ON INDUSTRIAL INFORMATICS, VOL. X, NO. XX, DECEMBER 31, 2021}%
% {\MakeLowercase{\textit{et al.}}: }

\maketitle

\begin{abstract}
The advancement of Internet and Communication Technologies (ICTs) has led to the era of Industry 4.0. This shift is followed by healthcare industries creating the term Healthcare 4.0. In Healthcare 4.0, the use of IoT-enabled medical imaging devices for early disease detection has enabled medical practitioners to increase healthcare institutions' quality of service. However, Healthcare 4.0 is still lagging in Artificial Intelligence and big data compared to other Industry 4.0 due to data privacy concerns. In addition, institutions' diverse storage and computing capabilities restrict institutions from incorporating the same training model structure. This paper presents a secure multi-party computation-based ensemble federated learning with blockchain that enables heterogeneous models to collaboratively learn from healthcare institutions' data without violating users' privacy. Blockchain properties also allow the party to enjoy data integrity without trust in a centralized server while also providing each healthcare institution with auditability and version control capability. 
\end{abstract}


% \fi
\IEEEpeerreviewmaketitle

\section{Introduction}
Ubiquitous computing, such as Artificial Intelligence (AI), the Internet of Things (IoT), and data mining, has transformed the manufacturing and engineering sectors, introducing the digitized Industrial era, also known as Industry 4.0 (I4.0). With the advent of I4.0, organizations have incorporated information and communication technologies (ICTs) to provide more efficient, scalable, and flexible services. I4.0 in the healthcare sector introduces the term Healthcare 4.0 (H4.0). The adoption of H4.0 in the healthcare system is argued to enable the shift from hospital-centered to patient-centered services, in which the interconnected healthcare ICTs are personalized based on the patients' needs and integrated to produce the best patient health outcome \cite{pace2018edge}. 

Despite the benefit, H4.0 is still lagging in Artificial Intelligence (AI) and big data compared with other sectors in I4.0. One constraint is that H4.0 usually incorporates clients' sensitive information. Hence, data sharing for AI model training may be constrained, resulting in insufficient data representation \cite{kaissis2021end}. Thus, trained AI models may perform poorly. 

The Federated Learning (FL) method was proposed to overcome data privacy concerns. FL allows multiple parties to train a single global model on a centralized server using their own local data without sharing the data. This is done by sharing server model parameters with each participant. Although adherence to regulations and data privacy are enhanced using the proposed method, FL still suffers several aspects. In defining the problem, we focus on privacy preservation on the machine learning model for image classification task using FL in H4.0.

\begin{figure}[!tbh]
\centering
\includegraphics[width=1\linewidth]{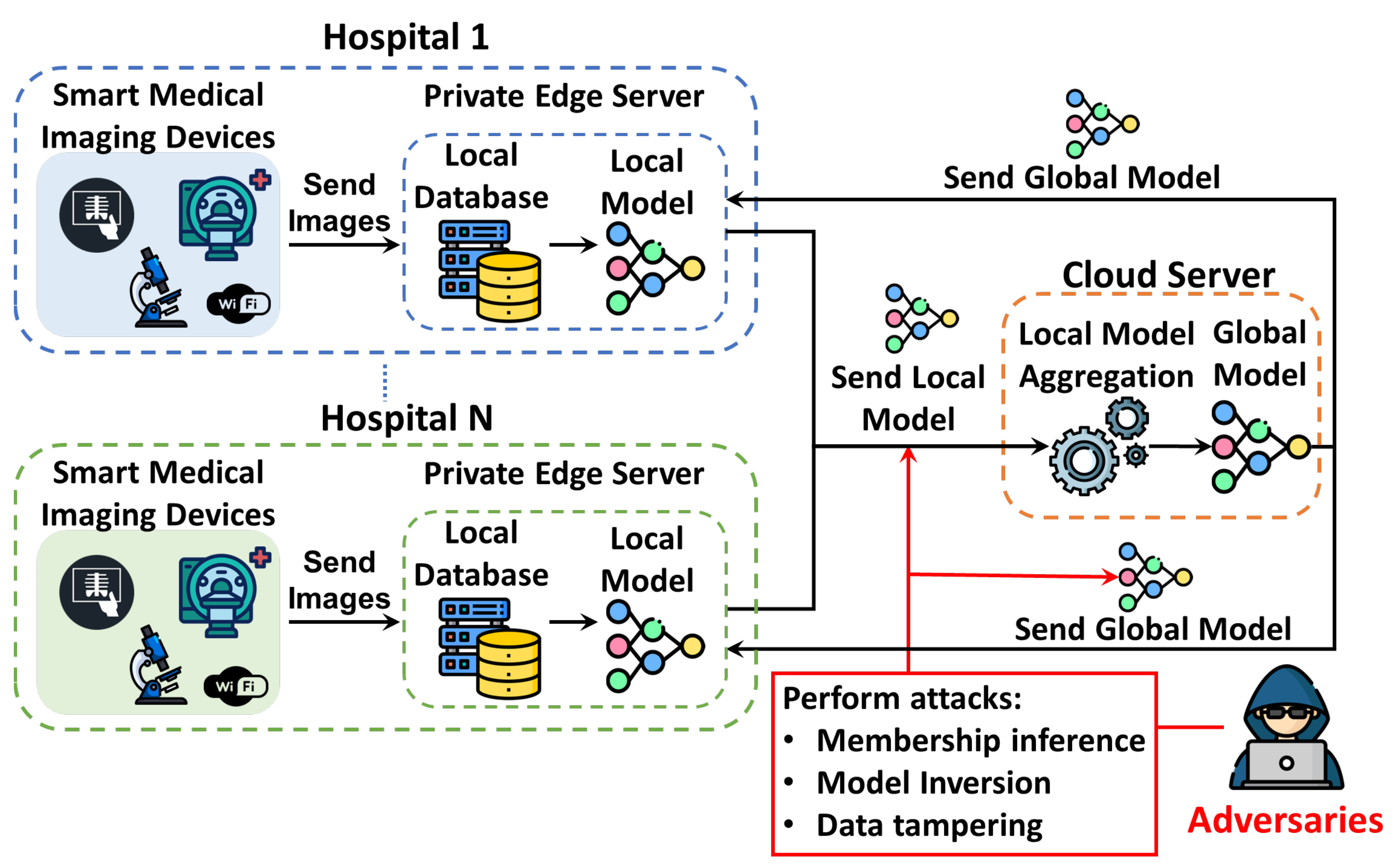}
\caption{Centralized FL without privacy-preservation method}
\label{fig: probStatement}
\end{figure}

As shown in Figure \ref{fig: probStatement}, hospitals participating in the FL process first collect medical data from their patients and store it locally in their local database. These data are kept secret from other hospitals and external parties to ensure patients' privacy. Then, the data are fed to the corresponding hospitals' local model to enhance its performance. Next, local models from each hospital are sent to the cloud server to be aggregated to create a global model. Eventually, the global model parameters are sent back to the hospitals for the subsequent FL process.

Intuitively, FL is safer than centralized training because data collected by each hospital are kept secret by the corresponding party. However, current FL methods are still faced some practical challenges. Privacy has been an ongoing concern in FL. \cite{liu2020secure} argued that a trained local model contains rich semantic information that can be traced back, resulting in the reconstruction of raw data distribution used for training. \cite{song2020analyzing} and \cite{shokri2017membership} show a successful model inversion and membership inference attack that can determine whether a record was used as part of the machine learning model's training. Hence, if exposed to adversaries, a trained local model may be vulnerable to model inversion and membership inference attacks. Data integrity is another complexity in FL. The FL model parameters sent over the network are prone to tampering, resulting in incorrect model parameter training.

To tackle these issues, several existing studies have integrated blockchain with some privacy preservation methods, such as Differential Privacy (DP) \cite{zhao2020privacy, jia2021blockchain}, Homomorphic Encryption (HE) \cite{jia2021blockchain}, and Secure-Multiparty Computation (SMPC) \cite{shayan2020biscotti} in the FL scheme. Blockchain fool-proof resistance property enables the party to prevent data tampering, while the privacy-preserving method used in FL can prevent parties from disclosing clients' private information.

Nevertheless, the existing proposed methods consider FL schemes such as FedAVG \cite{sahu2018convergence} and FedSGD \cite{felbab2019optimization}, which assumes that all participants' devices have similar models. This may not be the case in practice since different machine learning model structures may be employed because of differences in edge devices' computing resources, power consumption, and storage capacity. Each institution may also have its policy, which states the machine learning model structure used in their system. Hence, when resources with heterogenous computing power are involved, efficiency is an additional practical constraint that adds to the complication of FL.

To tackle privacy issues caused by shared model parameters and to ensure shared model integrity, this paper proposes a privacy-preserving blockchain-based ensemble-integrated FL scheme for image classification tasks in the context of H4.0. In the proposed method, we assume that each hospital has a similar model structure. Hence, healthcare devices within the institution may perform FL using the existing FL algorithm. However, the learning model structure between different hospitals may differ. For this, we propose a weighted ensemble Deep Learning (DL) to enable the aggregation of heterogeneous model structures to produce a final global model. We use model accuracy evaluation to determine the ensemble model weights. This allows misbehaving or lower-performing models to contribute less to the final outcome. In this manuscript, we consider each entity to be honest but curious. Hence, we utilize an SMPC-based method for ensemble model evaluation across hospitals to ensure privacy guarantees of the models produced by hospitals. Finally, to ensure data integrity and auditability, we leverage the use of blockchain.

\section{Related Work}\label{sec:related}

In H4.0, the incorporation of ICT devices and AI poses great challenges to privacy protection and data integrity in real-world applications. FL was proposed in \cite{mcmahan2016federated} to jointly train a global model without sharing the local datasets with the global server. Intuitively, a basic privacy guarantee can be achieved by this method because the private datasets are not transmitted to the global server. However, FL alone is not sufficient to provide a privacy guarantee. This has been proven in \cite{shokri2017membership}, and \cite{fredrikson2015model}, where the authors have demonstrated successful membership inference and model inversion attacks, respectively, on the exchanged FL model.

Previous studies have incorporated a DP method in the learning process in tackling these issues. For example, authors in \cite{phong2019privacy, zhang2019deeppar, wei2020federated, hu2020personalized} proposed a DP-based mechanism to obfuscate the trained local model parameters. Although DP integration can provide a better privacy guarantee to a certain extent, there is a trade-off between privacy and model accuracy. For this reason, SMPC schemes tailored to FL have been proposed. 

In \cite{ma2018deep}, the authors developed SMPC-based collaborative learning by combining  ElGamal encryption and Diffie-Hellman key exchange protocol to preserve data privacy and parameter privacy without sacrificing the resulting model's accuracy. \cite{9235504} further enhances the privacy preservation in collaborative learning by proposing SMPC-based collaborative learning that is resistant to generative adversarial networks. This is ensured by isolating participants from model parameters. Although both SMPC-based methods are able to produce a high-performing model, it requires a high cost in calculating complex functions. Therefore, implementing the SMPC scheme while each party uses the same machine learning model structure may not benefit the party with less computing power.

Additionally, FL does not provide tamper-proof attributes to ensure data integrity. The blockchain is a shared, immutable ledger where transactions are recorded in the blocks that are connected in chronological order. It has benefits in terms of data integrity, open autonomy, nontempering, and anonymous traceability \cite{9019859}.

The work in \cite{kang2019incentive} proposed an incentive-based mechanism in blockchain for robust FL model updates. Specifically, the blockchain is used to store each participant's 'reputation' score based on their performance history. The downside of the work is that for any updates that are not classified as malicious, clients will be positively rewarded. This is also true for clients whose updates are regarded as malicious. Similarly, authors in \cite{8894364} presented an incentive-driven mechanism in blockchain for FL called DeepChain. DeepChain aims to encourage parties to participate actively and behave correctly in FL training by giving rewards for their contributions. Nevertheless, the cost given for updates of each participant has yet to be considered. Hence, further investigation of the system costs and rewards profit needs to be done to ensure that clients and model owners benefit from the system. 

While considering the privacy-preservation on the local parameter updates, the work \cite{arachchige2020trustworthy, 9684698} presented a joint framework of blockchain, DP, and FL to protect data privacy in Industrial Internet-of-Things (IIoTs). A DP approach is applied by employing a randomized mechanism during the local model training, producing a differentially private model update to minimize individual record identification. Smart contracts are used for parameter exchange between the participants and the central authority to provide transparency. Thus, it enhances the reliability and safety of the FL process against external adversaries. Nonetheless, a privacy-preservation method such as DP provides a high privacy guarantee at the cost of model accuracy. 

Author in \cite{9763363} proposed a secure aggregation method using  Intel Software Guard Extension (SGX)-based Trusted Execution Environment (TEE) to securely aggregate local models in IIoTs. The proposed method is able to preserve the privacy of local model parameters. However, unlike the work proposed in \cite{kang2019incentive} the proposed method does not consider local model evaluation before the aggregation. Therefore, malicious updates are considered to have the same contribution to the global model.

Unlike the previous studies, our work does not utilize an incentive-based mechanism. Our proposed method focuses on a general model evaluation for a fair model contribution. Each participant can evaluate other participants' machine learning models to determine how well they perform towards unforeseen data. Then, based on the evaluation, each participant's machine learning models are weighted to determine their contribution to the final predictions. The evaluation process is handled by the blockchain nodes and recorded on tamper-proof storage. Furthermore, we utilized the SMPC privacy preservation method to preserve the machine learning model privacy while considering that each participant has different computing power. Therefore, we proposed a hierarchical ensemble federated learning, which allows participants to define their own model structure.

\section{Methodology}\label{sec:methodology}

This section first presents the architecture overview of the proposed method. Then, we discuss each process: federated learning, encrypted inference for ensemble model evaluation, and blockchain for data integrity and trustworthiness. The summary of notations used in the methodology can be seen in Table \ref{tab: notation}.

\subsection{Architecture Overview}

\begin{figure*}[tbh!]
\centering
\includegraphics[width=0.9\linewidth]{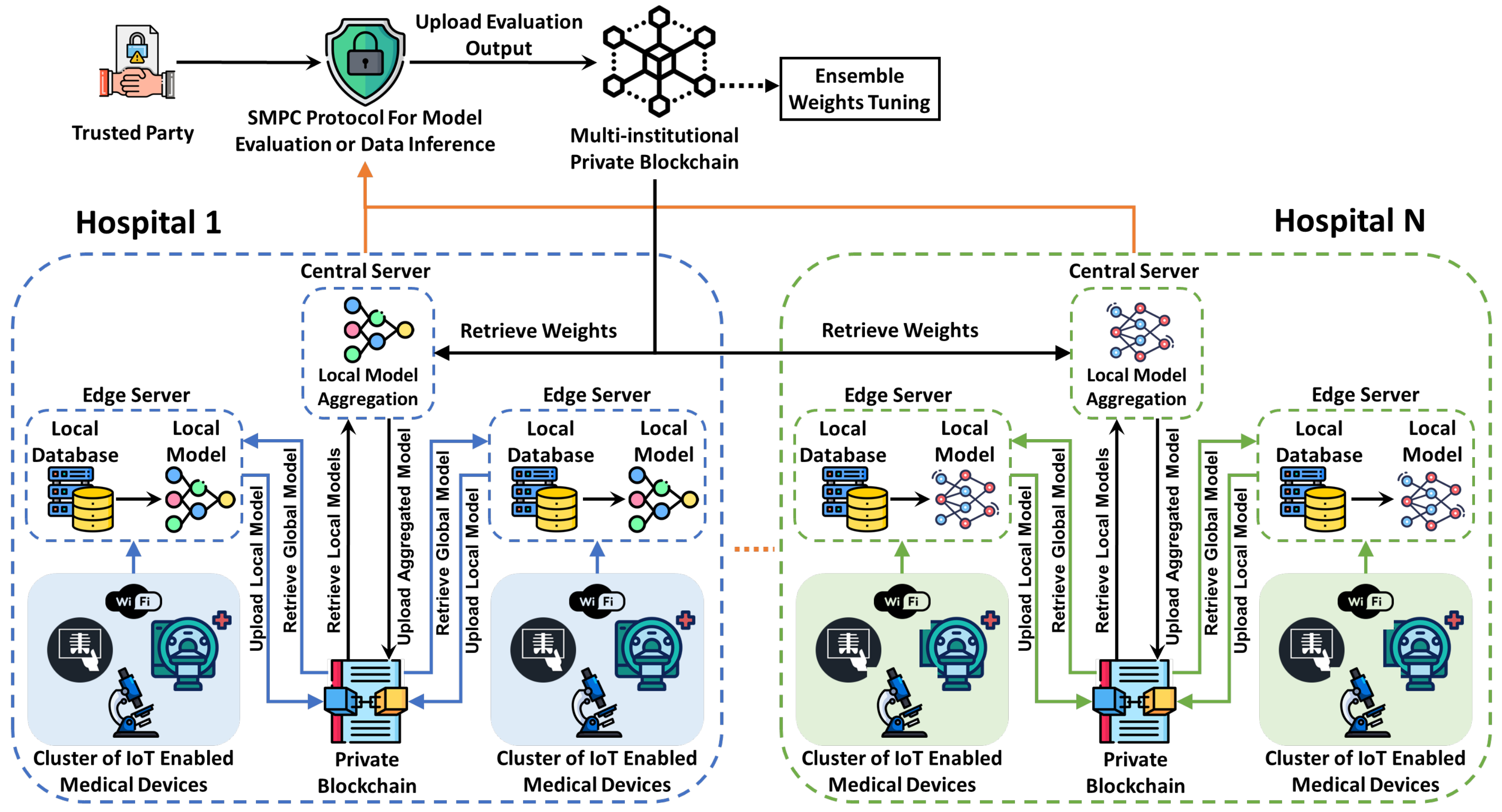}
\caption{Overview of the proposed architecture}
\label{fig: ArchOverview}
\end{figure*}

Our proposed architecture comprises edge servers, central servers, hospitals, private blockchain, private blockchain for multi-institutions, and a Trusted Third Party (TTP). In tackling the constraints in existing studies of FL, we proposed the architecture shown in Figure \ref{fig: ArchOverview}. We consider that there are multiple hospitals $\mathcal{H}_i \in \mathcal{H}$ that have their policy on the DL model structure to be used in their system. $\mathcal{H}$ are considered as honest-but-curious entities. Each $\mathcal{H}$ has edge servers  $E_{i} \in E$ which are connected to a cluster of IoT-enabled or smart medical imaging devices $\mathcal{C}_i$. Edge servers from the same hospital are considered to have the same computing power. Hence, a hospital must apply the same learning model structure to all of the edge servers based on its policy. However, edge servers from different hospitals may use different model structures based on the affiliated hospital's policy.

The learning process of a hospital in our architecture starts with $E_{i} \in E$ training their local model $\mathcal{M}_{i}$ using the image data collected from $\mathcal{C}_i$. The trained local model is then verified and stored in a private blockchain $\mathcal{B}_i$ owned by $\mathcal{H}_i$. Then, $\mathcal{H}_i$ collects all local models from $\mathcal{B}_i$ to be aggregated in their central server $\mathcal{S}_i$ to create a global model $\mathcal{GM}_i$. For each transaction, $\mathcal{H}_i$ creates a smart contract maintained within their private blockchain $\mathcal{B}_i$. The private blockchain enables the hospital to provide local data integrity. 

For hospitals to collaborate, an SMPC protocol is followed to perform encrypted inference, in which output is intended solely to infer the data or to be used further for model evaluation using other $\mathcal{H}_i$ data. The SMPC protocol is assisted by a Trusted Third Party (TTP) in providing necessary variables to keep the computation secret. Since SMPC is used during this process, only shares of data and models are exchanged between hospitals for their model evaluation. Therefore, privacy is preserved since the actual value of the hospital's model parameters and their data are kept secret. From the SMPC process, $\mathcal{GM}_i$ produces classification probabilities of data provided by all $\mathcal{H}$ for evaluation. These probabilities are then sent to the multi-institution private blockchain network $BM$. Each node of the blockchain $BM$ will then perform the ensemble weight tuning calculation and verify the value of the outcome. When verified, the fine-tuned weights are recorded in the tamper-proof storage of $BM$.

% \vspace{-18mm}
\begin{table}[!tbh]
\fontsize{10}{14}\selectfont
\begin{center}
\caption{Notations}
\label{tab: notation}
\scalebox{0.8}{
{
\begin{tabularx}{\linewidth}{
    >{\hsize=.3\hsize}X
    >{\hsize=.6\hsize}X
}
\toprule
    $\mathcal{H}$ & Set of hospitals\\
    $\mathcal{H}_i$ &  $\mathcal{H}_i \in \mathcal{H}$\\
    $E$ & Set of edge servers \\
    $E_i$ & $E_i \in E$ \\
    $C_i$ & Cluster of smart medical imaging devices \\
    $\mathcal{M}_i$ & Local model generated by $E_i$ \\
    $B$ & Set of private blockchains owned by $\mathcal{H}$ \\
    $B_i$ & $B_i \in B$ \\
    $S_i$ & Central server owned by $H_i$ \\
    $\mathcal{GM}_i$ & Global model produced by $H_i$ \\
    $BM$ & Multi-institution private blockchain \\
    $T$ & Total number of communication rounds \\
    $e$ & Epochs \\
    $k$ & Number of participants in federated learning \\
    $t$ & Communication round \\
    $J$ & Set of randomly chosen federated learning participants from $E$ \\
    $j$ & $j \in J$ \\
    $w_t$ & Weights of $\mathcal{GM}_i$ in the current $t$ \\
    $w_e^j$ & Weights of $\mathcal{M}_i$ in the current $e$ \\
    $\eta$ & Learning rate \\
    $g_j$ & Gradient calculated during $\mathcal{M}_i$ training \\
    $N_j$ & Current total number of samples used by $j$ for training \\
    $\theta_j$ & Current state of the model parameters \\
    $\nabla$ & Derivatives with respect to $\theta_j$ \\
    $x_i$ & Input data \\
    $y_i$ & $x_i$ true label \\
    $f(x_i)$ & Prediction of $x_i$ \\
    $l(\cdot, \cdot)$ & Loss function \\
    $N$ & Total number of samples used by $J$ \\
    $[\cdot]$ & Arithmetic secret share value \\
    $<\cdot>$ & Binary secret share value \\
    $P_{\mathcal{H}_i}$ & Output probabilities of $\mathcal{GM}_i$ \\
    $h_i$ & Hash value to the corresponding $\mathcal{GM}_i$ or $\mathcal{M}_i$ \\
    $\tilde {P}()$ & Prediction of ensemble learning model \\
    ${\alpha}$ & Sets of ensemble weights \\
    $\alpha$ & $\alpha \subset {\alpha}$ \\
    $\alpha_i$ & An ensemble weight value \\
    $\alpha_b$ & Best set of ensemble weights combination \\
    $W$ & A list of random value, $0 \leq W \leq 1$ \\
\bottomrule
\end{tabularx}
}
}
\end{center}
\end{table}

\subsection{Federated Learning}

For model aggregation in the proposed architecture, FedAVG \cite{sahu2018convergence} is used. Algorithm \ref{alg: fedAVG} shows the steps of FedAVG in detail. 

\SetAlFnt{\small}
\begin{algorithm}[ht!]
\SetAlgoNoLine
\caption{FedAVG}
\label{alg: fedAVG}
\KwIn
{
    \begin{minipage}[t]{10cm}%
     \strut
        $w_t$ - weights of $\mathcal{GM}_i$ owned by $\mathcal{H}_i$
     \strut
    \end{minipage}%
}
\KwOut{
    \begin{minipage}[t]{8cm}%
     \strut
        $Agg\mathcal{M}_j$ - Aggregated local models
     \strut
    \end{minipage}%
}
    \textbf{Initialization:}
    
    communication rounds, $T$
    
    epochs, $e$
    
    number of participants, $k$
    
    \textbf{begin}
    
    \For{$t \leftarrow 0$ \textbf{to} $T$}
        {
        
            $J \leftarrow random\_participant(E, k)$
            
            \For{$j \in J$}{
                
                recieve $w_t$ from $S_i$
                
                $w_{e}^j \leftarrow w_t$
                
                \For{$e \leftarrow 0$ \textbf{to} $e$}
                {
                
                    $w_{e+1}^j \leftarrow w_{e}^j - \eta g_j$
                
                }
            }
            
            $Agg\mathcal{M}_j \leftarrow \sum^{J}_{j = 1} \frac{N_j}{N} w_{e+1}^j$
            
        }

    \textbf{return} $Agg\mathcal{M}_j$
    
    \textbf{end}

\end{algorithm}

The process starts with central server defining the total number of communication round $T$,  epochs $e$, and number of participants $k$. Then, for each communication round $t$ the server randomly chooses $k$ participants from $E$. The pool of randomly chosen participants is denoted as $J$. Next, server sends $w_t$, which represents the $\mathcal{GM}_i$ weights, to each participant $j \in J$. Each $j$ then copy $w_t$ to their local model variable $w_{e}^j$. Next, they train the model and obtain the updated weight $w_{e+1}^j$ by calculating weights update function $ w_{e}^j - \eta g_j$. Here, $w_{e}^j$ represents the current weight, $\eta$ represents the learning rate, and $g_j$ is the gradient calculated during the training. The gradient can be calculated as shown in Equation \ref{eq: grad}
\begin{equation}
    g_{j} = \frac{1}{N_j} \sum^{N}_{i = 1} \nabla_{\theta_{j}}l(f(x_{i}), y_{i}),
    \label{eq: grad}
\end{equation}
where, $g_{j}$ is the gradient of the current step, $N_j$ is the number of samples used in the current training step by participant $j$, $\theta_{j}$ is the current state of the model parameter, $\nabla$ is used to refer to the derivative with respect to every parameter, $f(x_{i})$ is the model prediction with respect to input sample $x_{i}$, $y_{i}$ is the true label of input sample $x_{i}$, and $l()$ is the loss function.

When model training by each participant is done, they send $w_{e+1}^j$ to the server. The server then aggregates the received local model producing an aggregated local model $Agg\mathcal{M}_j$ using the equation as follows:

\begin{equation}
    \sum^{J}_{j = 1} \frac{N_j}{N} w_{e+1}^j
    \label{eq: avgAgg}
\end{equation}

Here, $j$ is a specific participant within the training phase, $N_j$ is the number of $j$ participant's training samples, and $N$ is the total samples used for training across all participants. The resulting aggregated model will then replace the current $\mathcal{GM}_i$.

\subsection{Encrypted Inference for Ensemble Model Evaluation}

Each hospital can enjoy the ensemble model by combining the output probabilities from multiple $\mathcal{GM}_i$. However, since the hospitals are assumed as honest-but-curious entities, sharing $\mathcal{GM}_i$ to other $\mathcal{H}$ without privacy is not possible. Hence, an SMPC protocol is used in securing the $\mathcal{GM}_i$. We consider using arithmetic and binary secret sharing to implement secure computations for data inference.

In arithmetic secret sharing an input value $x \in \mathbb{Z}/Q\mathbb{Z}$ is shared amongst $\mathcal{H}$. Here $\mathbb{Z}/Q\mathbb{Z}$ is a ring with $Q$ elements. To share the value $x$, $\mathcal{H}$ generate a pseudorandom zero-share \cite{cramer2005share}. The secret shares from a value $x$ is denoted as $[x] = \{[x]_{\mathcal{H}_i}\}_{\mathcal{H}_i \in \mathcal{H}}$, where $[x]_{\mathcal{H}_i} \in \mathbb{Z}/Q\mathbb{Z}$ is
$x$ share of party $\mathcal{H}_i$. The shares must fulfill a condition such that the sum of all shares reconstructs the value of $x$ as shown in Equation \ref{eq: sumShare}.
\begin{equation}
    x = \sum_{\mathcal{H}_i \in \mathcal{H}} [x]_{\mathcal{H}_i}
    \label{eq: sumShare}
\end{equation}
In binary secret sharing, $x$ operates in $\mathbb{Z}/2\mathbb{Z}$. Shares of value $x$ can be denoted as $\left < x \right >$. All $\mathcal{H}$ shares must hold a condition such that $x = \oplus_{\mathcal{H}_i \in \mathcal{H}} \left < x \right >_{\mathcal{H}_i}$. 

Since both binary and arithmetic secret sharing have homomorphic properties, they can be used for secure computation. Operations required in our model include addition, multiplication, and comparison. Private addition and multiplication can be done under arithmetic secret sharing, while comparison falls under binary secret sharing. Hence, secret shares conversions from $[x]$ to $\left < x \right >$ and vice versa are needed. The conversion of $[x]$ to $\left < x \right >$ is done by creating binary secret share of every bits in  $[x]_{\mathcal{H}_i}$ such that it satisfy $\left < x \right > = \sum_{\mathcal{H}_{i} \in \mathcal{H}} \left <[x] \right >$. To convert $\left < x \right >$ to $[x]$, the equation $[x] = \sum^{B}_{b=1} 2^b \left [ \left < x\right >^{(b)} \right ]$ is used. Here, $B$ is the total number of bits in $\left < x \right >$ and $b$ represents the $b$-th bits of binary share $\left < x \right >$. To calculate $\left [ \left < x\right >^{(b)} \right ]$, a TTP generates $\left ( \left [ r^{(b)} \right ],  \left < r^{(b)} \right > \right )$. Next, $ \left [ \left < x\right >^{(b)} \right ] = \left [ r^{(b)} \right ] + z^{(b)} - 2 \left [ r^{(b)} \right ] z^{(b)}$ is calculated. Here, $z^{(b)}$ is obtained by masking $\left < x\right >^{(b)}$ with $\left < r^{(b)} \right >$.

In private addition, $\mathcal{H}_i \in \mathcal{H}$ adds their shares such that $[z]_{\mathcal{H}_i} = [x]_{\mathcal{H}_i} + [y]_{\mathcal{H}_i}$. For multiplication, random Beaver triples proposed by \cite{beaver1991efficient} is implemented. In the process a TTP provides triples $([a], [b], [c])$, such that $c = ab$. Each $\mathcal{H}_i \in \mathcal{H}$ then compute $[\epsilon] = [x] - [a]$ and $[\delta] = [y] - [b]$. Value $[\epsilon]$ and $[\delta]$ are then decrypted producing $\epsilon$ and $\delta$. Finally, $[x][y] = [c]+\epsilon[b]+[a]\delta+\epsilon \delta$ is calculated. For comparison, an evaluation function $[z < 0]$ is used. To securely compute comparison of an arithmetic share, first, $[z]$ is converted into $\left < z \right >$. Then, a sign bit is computed using $\left < b \right > = \left < z \right > \gg (L-1)$, where $L$ is the length of bits. Finally, the resulting bit is converted back into arithmetic sharing $[b]$. When checking if a value is greater than 0, for example ReLU activation function, the function can be written as $ReLU([x]) = [x][x < 0]$. On the other hand, when comparing two values, the two shares are subtracted $[z] = [x] - [y]$, then it is evaluated using $[z < 0]$.

The process of encrypted model evaluation is shown in Algorithm \ref{alg: MO} and \ref{alg: DO}. Algorithm \ref{alg: MO} shows the steps done on the Model Owner (MO) site. Assume that $\mathcal{H}_i$ is an MO. Suppose that $w_t$ is the parameter of $\mathcal{GM}_i$ and there exists $h$ number of hospitals in the system. The MO first creates $h$ shares of $w_t$. The shares $\{[w]_{\mathcal{H}_i}\}_{\mathcal{H}_i \in \mathcal{H}}$ is then send to the respective hospital. Next, $\mathcal{H}_i$ receives shares of input data and their labels $([x], y)$ from other hospitals to evaluate the model. In this case, we assume that $\mathcal{H}_i \in \mathcal{H}$ other than MO acts as DO. Next, MO starts secure computation across all participants to produce shares of output probabilities $\left [ P_{\mathcal{H}_i} \right ]$. Finally, $\left [ P_{\mathcal{H}_i} \right ]$ is decrypted using Equation \ref{eq: sumShare}. The decrypted probabilities $ P_{\mathcal{H}_i}$ and corresponding labels $y$ are then sent to the cloud for ensemble model weights fine-tuning.

\SetAlFnt{\small}
\begin{algorithm}[ht!]
\SetAlgoNoLine
\caption{Encrypted Inference (MO)}
\label{alg: MO}
\KwIn
{
    \begin{minipage}[t]{10cm}%
     \strut
        $w_t$ - weights of $\mathcal{GM}_i$ owned by $\mathcal{H}_i$
        
        $h$ - number of participating $\mathcal{H}$
     \strut
    \end{minipage}%
}
\KwOut{
    \begin{minipage}[t]{8cm}%
     \strut
        $P_{\mathcal{H}_i}$ - output probabilities
     \strut
    \end{minipage}%
}
    \textbf{Initialization:}
    
    Number of $\mathcal{H}$, $h$
    
    \textbf{begin}
    
    $\{[w_t]_{\mathcal{H}_i}\}_{\mathcal{H}_i \in \mathcal{H}} = create\_share(w_t, h)$
    
    \ForEach{$\mathcal{H}_i \in \mathcal{H}$}{
    
        $send([w_t]_{\mathcal{H}_i}, \mathcal{H}_i)$
    
    }
    
    Recieve $([x], y)$ from DOs
    
    $[P_{\mathcal{H}_i}] = start\_secure\_computation([w_t]_{\mathcal{H}_i}, [x])$
    
    $P_{\mathcal{H}_i} = decrypt([P_{\mathcal{H}_i}])$
    
    \textbf{return} $P_{\mathcal{H}_i}, y$

\end{algorithm}

Algorithm \ref{alg: DO} shows the steps required at Data Owner (DO) site. First, the DO prepares pre-processed data for evaluation $(x, y)$. Here $x$ denotes image data, and $y$ is the true label of the data. DO then creates $h$ shares of $x$, which is denoted as $\{[x]_{\mathcal{H}_i}\}_{\mathcal{H}_i \in \mathcal{H}}$. The shares are then sent to each $\mathcal{H}_i \in \mathcal{H}$. Finally, it starts the secure computation process using the weight received from MO to produce $[P_{\mathcal{H}_i}]$, which is then sent back to the MO to be decrypted.

\SetAlFnt{\small}
\begin{algorithm}[ht!]
\SetAlgoNoLine
\caption{Encrypted Inference (DO)}
\label{alg: DO}
\KwIn
{
    \begin{minipage}[t]{10cm}%
     \strut
        $[w]_{\mathcal{H}_i}$ - shares of model parameters
     \strut
    \end{minipage}%
}
\KwOut{
    \begin{minipage}[t]{8cm}%
     \strut
        $[P_{\mathcal{H}_i}]$ - output probabilities shares
     \strut
    \end{minipage}%
}
    \textbf{Initialization:}
    
    Input data and labels for model evaluation, $(x, y)$
    
    \textbf{begin}
    
    $\{[x]_{\mathcal{H}_i}\}_{\mathcal{H}_i \in \mathcal{H}} = create\_share(x, h)$
    
    \ForEach{$\mathcal{H}_i \in \mathcal{H}$}{
    
        $send([x]_{\mathcal{H}_i}, \mathcal{H}_i)$
    
    }
    
    $[P_{\mathcal{H}_i}] = start\_secure\_computation([w]_{\mathcal{H}_i}, [x])$
    
    \textbf{return} $[P_{\mathcal{H}_i}]$

\end{algorithm}

\subsection{Blockchain for Data Integrity and Trustworthiness}

Blockchain in the proposed architecture is divided into two categories. The private blockchain $B_i$ that is owned by $\mathcal{H}_i \in \mathcal{H}$ and a multi-institutional private blockchain $BM$ which can be accessed by all of the $\mathcal{H}$. 

\begin{figure}[!tbh]
\centering
\includegraphics[width=0.8\linewidth]{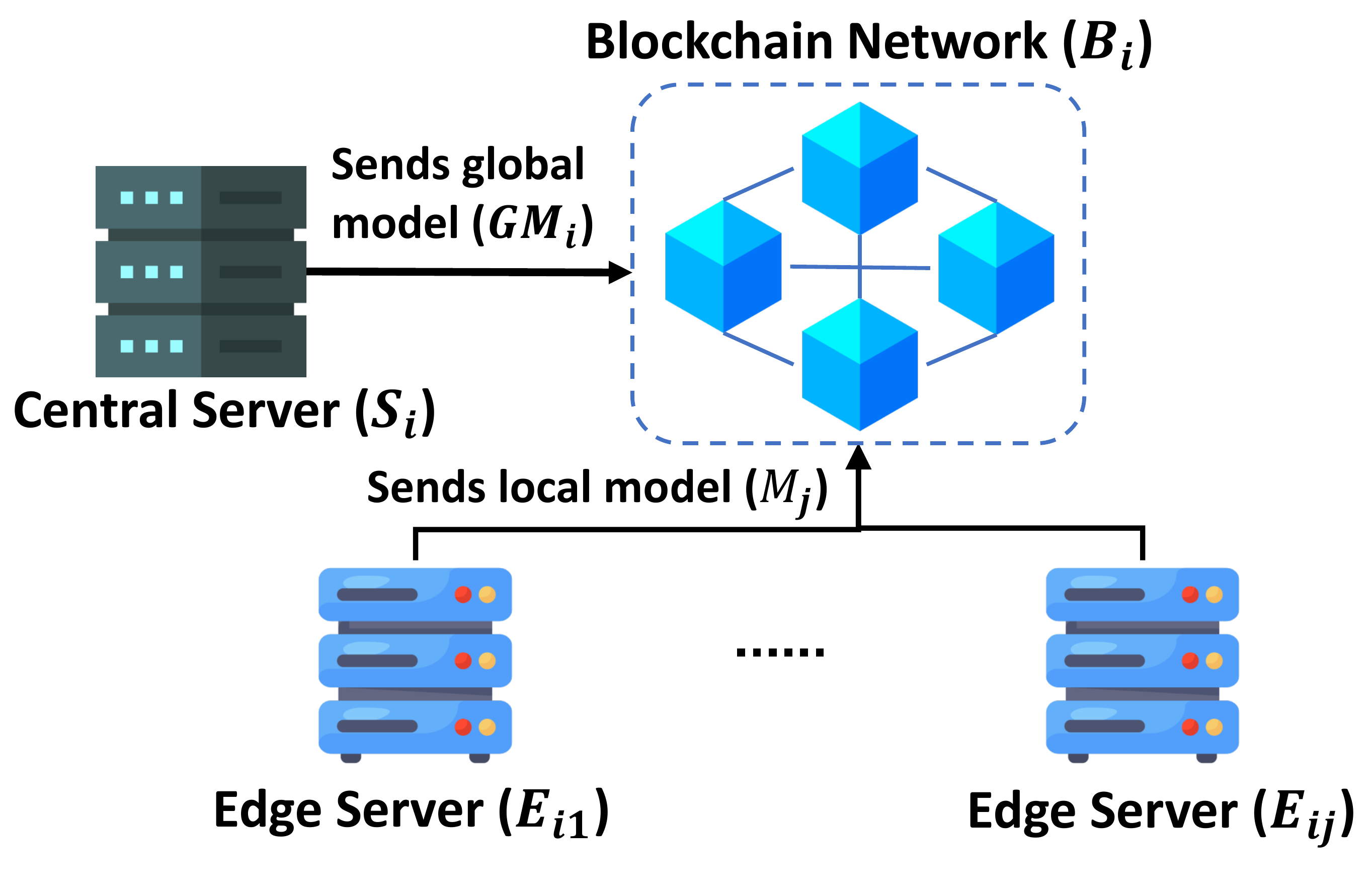}
\caption{Blockchain for model verification}
\label{fig: blockchainVer}
\end{figure}

$B_i$ ensures that each local and global model update is verifiable and trustworthy. Meaning that the internal party will not be able to tamper parameters of the model being exchanged. In the process, as shown in Figure \ref{fig: blockchainVer}, a private blockchain network $B_i$ receives local model updates $\mathcal{M}_i$ from edge server $E_{i}$ or a global model $\mathcal{GM}_i$. Then, $B_i$ generates the hash $h_i$ of the respective model and the model ID $\mathcal{M}_{id}$. Next, the miner within $B_i$ join together to run a consensus mechanism to verify the respective model.

\begin{figure}[!tbh]
\centering
\includegraphics[width=1\linewidth]{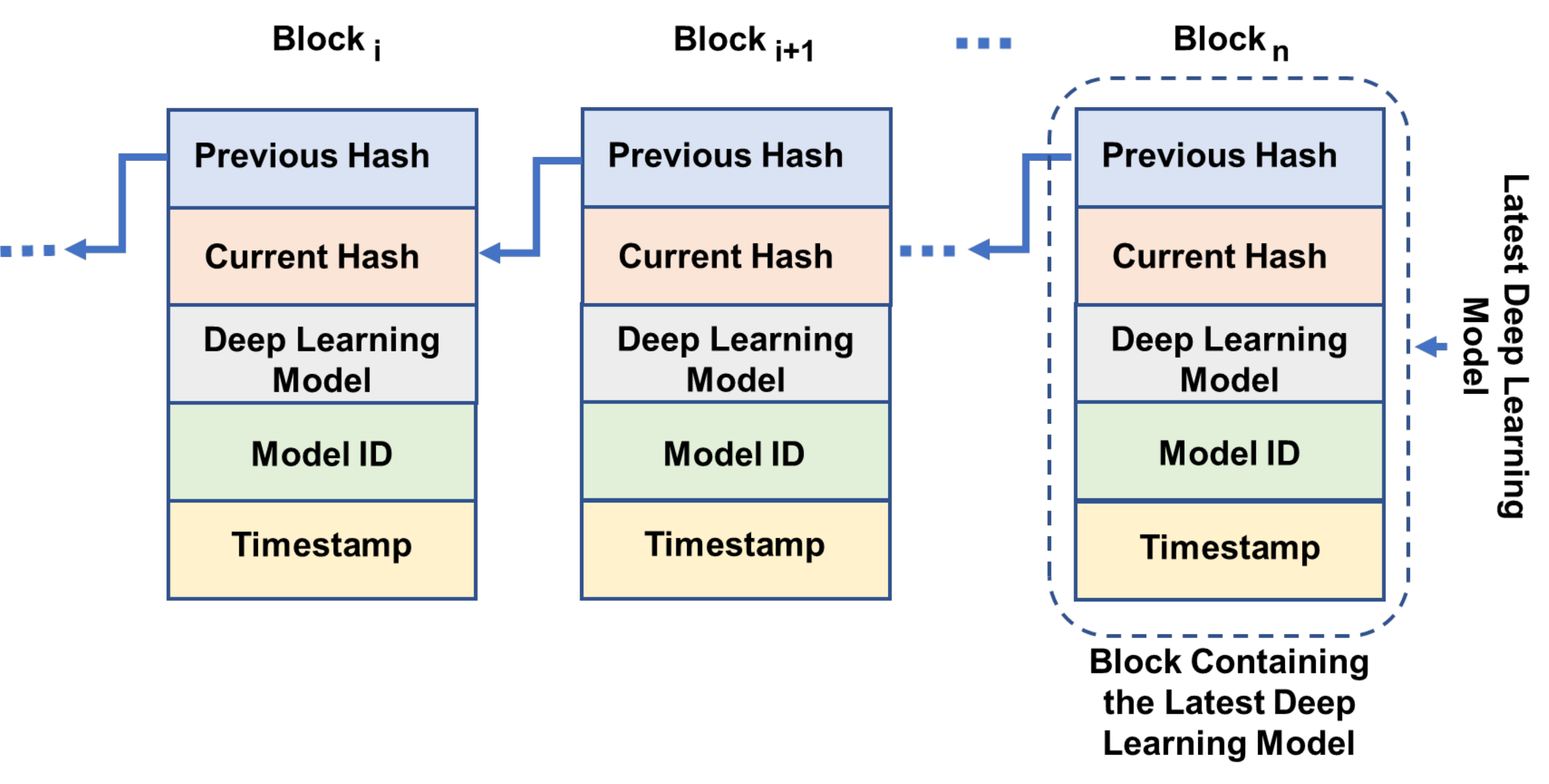}
\caption{Private blockchain data structure}
\label{fig: priv}
\end{figure}

When the majority of blockchain nodes in $B_i$ produce the same hash of the corresponding model and $\mathcal{M}_{id}$, a block is appended to the blockchain block containing the model and its hash. The consensus mechanism for model verification can be seen in Algorithm \ref{alg: privBlockVer}, while the blockchain data structure can be seen in Figure \ref{fig: priv}.

\SetAlFnt{\small}
\begin{algorithm}[ht!]
\SetAlgoNoLine
\caption{Private Blockchain Model Verification}
\label{alg: privBlockVer}
\KwIn
{
    \begin{minipage}[t]{10cm}%
     \strut
        $\mathcal{M}_i$ - local model to be verified
        
        $\mathcal{GM}_i$ - global model to be verified
     \strut
    \end{minipage}%
}
\KwOut{
    \begin{minipage}[t]{8cm}%
     \strut
        $h_i$ - hash of the model and model id
     \strut
    \end{minipage}%
}

    \textbf{Initialization:}
    
    Model hashes, $HM = \emptyset$
    
    Model id = $\mathcal{M}_{id}$
    
    \textbf{begin}
    
    \ForEach{$\mathcal{B}_i \in \mathcal{B}$}{
    
        $h_i = generate\_hash(\mathcal{M}_i||\mathcal{GM}_i, \mathcal{M}_{id})$
        
        $HM.add(h_i)$
    
    }
    
    add $\mathcal{M}_i||\mathcal{GM}_i$ and $\mathcal{M}_{id}$ to blockchain if $h_i \in HM$ are the same

\end{algorithm}

$BM$ in our scenario is responsible for ensemble model weights tuning and verification. The process is similar to $B_i$. However, instead of a model to be verified, each node in $BM$ is responsible for performing ensemble model weight tuning and verifying the best weight to be used for data inference.

\textbf{Ensemble Model Weights Tuning}. The weighted ensemble model enables us to combine multiple heterogeneous models to predict based on the proportion of their estimated evaluation. It also helps to reduce the number of prediction errors resulting in higher performance. In general, weighted ensemble learning can be denoted using Equation \ref{eq: weightedEns}.
\begin{equation}
     {\tilde {P}}(\mathbf {\alpha } )=\sum _{i=1}^{h}\alpha _{i}P_{\mathcal{H}_i}
    \label{eq: weightedEns}
\end{equation}
Here, $\tilde {P}()$ represents the predictions of the ensemble model, $\alpha$ is a list of weights, $h$ is the number of $\mathcal{H}$ participating, $\alpha_i$ is the weight assigned to $\mathcal{H}_i$, and $P_{\mathcal{H}_i}$ is the resulting prediction probabilities of input data $x$ on $\mathcal{GM}_i$. The best weight $\alpha_b$ is determined by calculating the number of correct predictions when experimenting with different values of $\alpha$. To find $\alpha_b$ we use Grid Search (See Algorithm \ref{alg: gridSearch}).

\SetAlFnt{\small}
\begin{algorithm}[ht!]
\SetAlgoNoLine
\caption{Weight tuning using Grid Search}
\label{alg: gridSearch}
\KwIn
{
    \begin{minipage}[t]{10cm}%
     \strut
     
        $\{P_{\mathcal{H}_i}\}_{\mathcal{H}_i \in {\mathcal{H}}}$ - output probabilities
        
        ${\{\mathcal{GM}_{id}\}}$ - a set of global model id
        
        $y$ - list of true label
     \strut
    \end{minipage}%
}
\KwOut{
    \begin{minipage}[t]{8cm}%
     \strut
        $\alpha_b$ - best weights combination
     \strut
    \end{minipage}%
}
    \textbf{Initialization:}
    
    list of possible weights, $\{W\}, 0 \leq W \leq 1$
    
    best weight, $\alpha_b = 0$
    
    best accuracy, $accuracy_b = 0$
    
    number of hospitals, $h$
    
    number of weights, $n$
    
    \textbf{begin}
    
    $\alpha = product(W, h)$
    
    \ForEach{${\alpha} \subset \{\alpha\}$}{
    
        ${\alpha} = l_1\_norm({\alpha})$
        
        \ForEach{$i \leftarrow 1$ \textbf{to} $n$}{
            
            ${\tilde {P}}(\mathbf {\alpha} )=\sum _{i=1}^{h}\alpha_{i}P_{\mathcal{H}_i}$
            
            $\tilde {P} = argmax(\tilde {P})$
            
            $accuracy = score(\tilde {P}, y)$
            
            \If{$accuracy > accuracy_b$}{
                
                $accuracy_b = accuracy$
                
                $\alpha_b = \alpha$
                
            }
        
        }
    }
    
    Broadcast $\{\mathcal{GM}_{id}\}$ and $\alpha_b$ to blockchain network

\end{algorithm}

\begin{figure}[!tbh]
\centering
\includegraphics[width=1\linewidth]{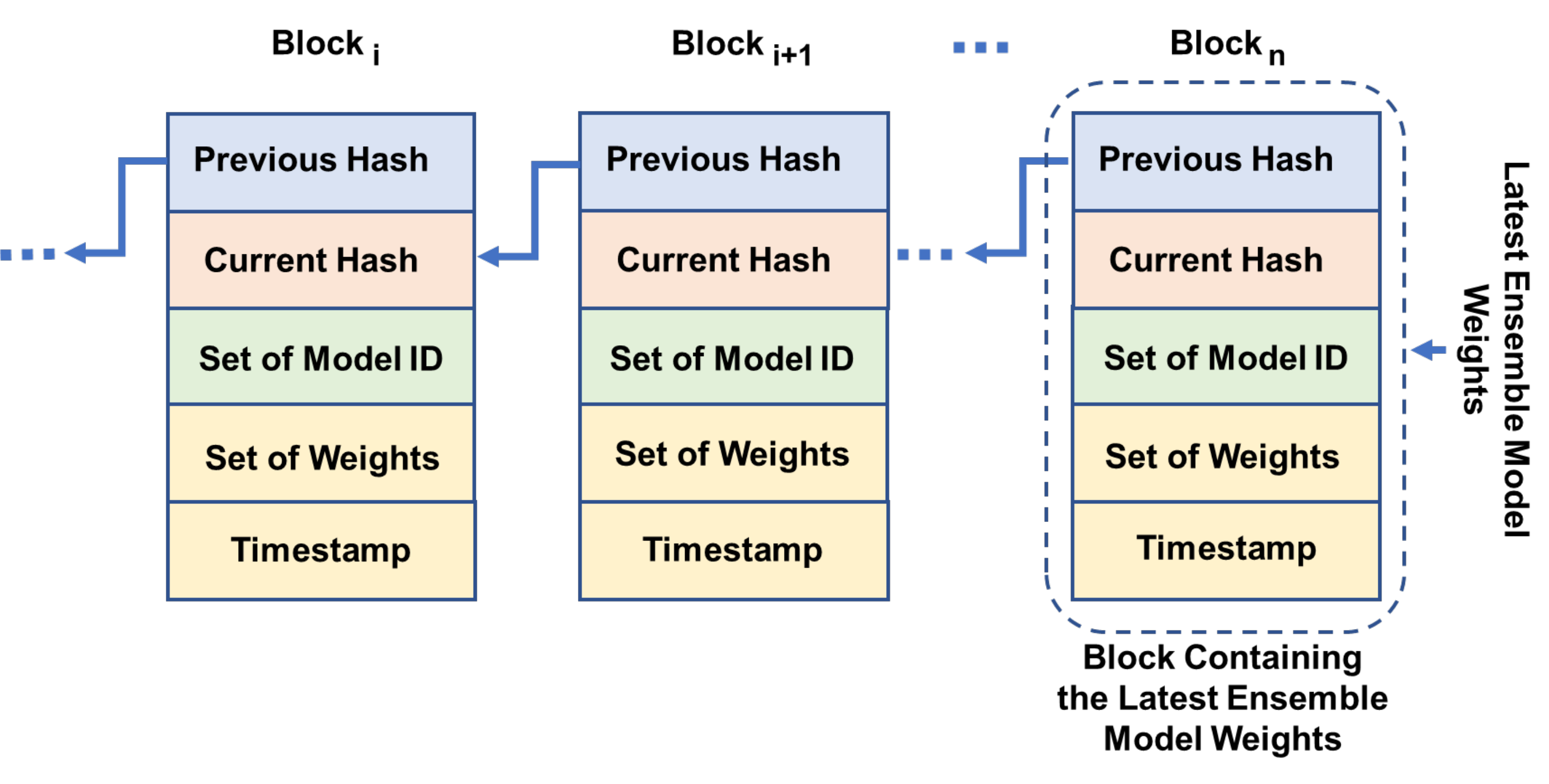}
\caption{Private multi-institutional blockchain data structure}
\label{fig: privMult}
\end{figure}

In the process, each node in $BM$ first receives output probabilities $\{P_{\mathcal{H}_i}\}_{\mathcal{H}_i \in {\mathcal{H}}}$ of global models $\mathcal{GM}_i \in \mathcal{GM}$ from all hospitals $\mathcal{H}$. All $\mathcal{GM}$ are tested against dataset in an orderly manner. Hence, the first probability output of $\{P_{\mathcal{H}_i}\}, \{P_{\mathcal{H}_{i+1}}\}, ..., \{P_{\mathcal{H}_n}\}$ refers to the output from the same data. It also receives a set of global model ID  $\{\mathcal{M}_{id}\}$ used to produce $\{P_{\mathcal{H}_i}\}_{\mathcal{H}_i \in {\mathcal{H}}}$. Then, all $BM_i \in BM$ creates a list of possible weights that is defined as $\{w_1, w_2, ..., w_n\} \in W, 0 \leq W \leq 1$, a variable to store a set of best weights $\alpha_b$, and a variable to keep track of the best accuracy $accuracy_b$ obtained. Next, the cloud creates ${\alpha}$, which consists of Cartesian product of all weights combinations. The cartesian product is denoted as $W_1 \times W_2 \times ... \times W_n$. A sample subset $\alpha \in \{ \alpha \}$ can consists of combination weights ${w_1, w_2, ..., w_1}$ with each subset consists of $n$ number of weights. Each weight in $\alpha$ represents the contribution proportion of $P_{\mathcal{H}_i}$. For each weights combination $\alpha \in \{ \alpha \}$, an $l_1 normalization$ technique is used to calculate ${\tilde {P}}(\mathbf {\alpha } )$ using equation \ref{eq: weightedEns}. Then, a prediction ${\tilde {P}}$ is produced by taking prediction with the highest probability value. Finally, an accuracy score is calculated by comparing ${\tilde {P}}$ with the true label $y$. If the current $accuracy$ is higher than the current best accuracy $accuracy_b$, then the current $\alpha$ is broadcasted to the blockchain network to be verified using consensus mechanism in Algorithm \ref{alg: ensWeightVer}. Data structure overview of the blockchain blocks can be seen in Figure \ref{fig: privMult}.

\SetAlFnt{\small}
\begin{algorithm}[ht!]
\SetAlgoNoLine
\caption{Ensemble Model Weights Verification}
\label{alg: ensWeightVer}
\KwIn
{
    \begin{minipage}[t]{10cm}%
     \strut
        $\alpha$ - set of model weights to be verified
        
        $\{\mathcal{GM}_{id}\}$ - sets of global model id
     \strut
    \end{minipage}%
}
\KwOut{
    \begin{minipage}[t]{8cm}%
     \strut
        $h_i$ - hash of the weights and model id
     \strut
    \end{minipage}%
}

    \textbf{Initialization:}
    
    Model hashes, $HM = \emptyset$
    
    \textbf{begin}
    
    \ForEach{$\mathcal{B}_i \in \mathcal{B}$}{
    
        $h_i = generate\_hash(\alpha, \mathcal{GM}_{id})$
        
        $HM.add(h_i)$
    
    }
    
    add $\alpha$ and $\mathcal{GM}_{id}$ to blockchain if $h_i \in HM$ are the same

\end{algorithm}

\section{Results and Discussion}\label{sec:exp}

This section discusses the testing environment, datasets, and the experimental setup used. We then compare the model accuracy of our proposed approach with the orthodox federated learning approach. Finally, we empirically measure the time consumption for ensemble weights tuning, encrypted inference, and blockchain smart contract execution. 

\subsection{Testing Environment}
We used AWS Sagemaker for our experiment. We chose AWS p3.2xlarge machines, which contain 1 Tesla V100 GPU with 16 GB GPU memory and 61 GB RAM. The experiments were carried out using Python version 3.7.

\subsection{Experimental Setup}
 In our experiment, we consider three hospitals participating in the privacy-preserving distributed learning setup. Each hospital has two participants participating in the federated learning process. Each hospital has a different pre-defined model structure. In our paper, we consider using AlexNet\cite{krizhevsky2017imagenet}, ResNet18\cite{szegedy2017inception}, and Net\cite{medmnistv2}.

\subsection{Datasets}
The effectiveness of the proposed model is tested against two medical image datasets. The training datasets and models are defined as follows:

\begin{itemize}
  \item \textbf{2D Colon Pathology}. This dataset consists of 3-channel RGB 28 $\times$ 28 2D colon pathology images from patients with colorectal cancer, which are classified into nine different categories. We retrieved the pre-processed images from \cite{yang2021medmnist}. The dataset consists of 89.996 training data, 10.004 testing data, and 7,180 validation data.
  
  \item \textbf{Breast Tumor}. This dataset is retrieved from \cite{spanhol2016breast}. It comprises 3-channel RGB 700 $\times$ 460 images of breast tumor tissue using different magnifying factors (40X, 100X, 200X, and 400X). The images consist of 9.109, with 2.480 classified as benign and 5.429 classified as malignant.
 
\end{itemize}

In our experiment, we divided the dataset into three partitions, namely, training, testing, and validation. For the 2D Colon Pathology, we follow the initial partition, while for the Breast Tumor dataset, we divided it into 7.000 training data, 1.000 testing data, and 1.109 validation data. The training dataset is used for federated learning within each of the hospitals. For this, the data is split evenly amongst all the participants within the hospitals. Validation data in our scenario is used only for ensemble weights tuning. Testing data is used to represent any unforeseen or future data to be predicted. This partition will be used to test our initial public model and the final ensemble model.

\subsection{CNN Model Configuration}

There are three CNN models that we use in this experiment. Two models are pre-trained models, AlexNet \cite{krizhevsky2014one} and ResNet18 \cite{he2016deep}. The other one is Net used in \cite{yang2021medmnist}. For simplicity, the input taken by all CNN models is set to be the same. That is an image with 3 $\times$ 100 $\times$ 100 in a format of color channel $\times$ height $\times$ width. The output of the CNN model is the probabilities of each class in the classification task. We also use the same settings for the three CNN models in terms of the training configuration. We set each model to have a learning rate which is set to 0.001, 20 epochs with a batch size of 128. 

\subsection{Ensemble-Federated Learning Model Accuracy}

We compare our work with three different setups to evaluate the effectiveness of our proposed method. The first one is the traditional centralized CNN model, where all data is collected in one database and used to train a single model. Here, we combine training and validation datasets for model training, while the testing data is used for the model evaluation. The second setup is centralized federated learning, which steps can be seen in Algorithm \ref{alg: fedAVG}. In the training process of federated learning, the training dataset is split evenly amongst participants, while the global model is tested against the testing dataset. The third setup is an FL with a TEE-based secure aggregation scheme. For this experiment, we also used the same experimental configuration used in the second setup. Our proposed method, the traditional FL, and TEE-based FL scheme, consider the use of three participants during the training. Table \ref{table: modelAcc} shows the accuracy results on the four setups.

\begin{table}[!th]
\small
\renewcommand{\arraystretch}{1.3}
\caption{CNN model accuracy in different setup}
\label{table: modelAcc}
\centering
\scalebox{0.8}{
    \begin{tabular}{|c|c|c|c|}
    \hline
    Dataset & Model Setup & Model Name & Accuracy \\
    \hline
    \multirow{10}{*}{\makecell{Colon\\Pathology}} & \multirow{3}{*}{Centralized} & $AlexNet$ & $88.56\%$ \\
    \cline{3-4}
    & & $ResNet18$ & $96.10\%$ \\ 
    \cline{3-4}
    & & $Net$ & $86.5\%$ \\ 
    \cline{2-4}
    & \multirow{3}{*}{FedAvg \cite{sahu2018convergence}} & $AlexNet$ & $84.70\%$ \\ 
    \cline{3-4}
    & & $ResNet18$ & $90.4\%$ \\ 
    \cline{3-4}
    & & $Net$ & $82.4\%$ \\ 
    \cline{2-4}
    & \multirow{3}{*}{TEE-based FL \cite{9763363}} & $AlexNet$ & $82.26\%$ \\ 
    \cline{3-4}
    & & $ResNet18$ & $87.5\%$ \\ 
    \cline{3-4}
    & & $Net$ & $80.08\%$ \\ 
    \cline{2-4}
    & \multirow{4}{*}{Ensemble-FedAvg} & $AlexNet + Net + ResNet18$ & $91.37\%$ \\
    \cline{3-4}
    & & $AlexNet^*$ & $86.14\%$ \\ 
    \cline{3-4}
    & & $ResNet18^*$ & $92.61\%$ \\ 
    \cline{3-4}
    & & $Net^*$ & $83.60\%$ \\ 
    \hline
    \multirow{10}{*}{\makecell{Breast\\Cancer}} & \multirow{3}{*}{Centralized} & $AlexNet$ & $85.05\%$ \\
    \cline{3-4}
    & & $ResNet18$ & $86.06\%$ \\ 
    \cline{3-4}
    & & $Net$ & $86.78\%$ \\ 
    \cline{2-4}
    & \multirow{3}{*}{FedAvg \cite{sahu2018convergence}} & $AlexNet$ & $83.75\%$ \\ 
    \cline{3-4}
    & & $ResNet18$ & $85.82\%$ \\ 
    \cline{3-4}
    & & $Net$ & $84.71\%$ \\ 
    \cline{2-4}
    & \multirow{3}{*}{TEE-based FL \cite{9763363}} & $AlexNet$ & $82.24\%$ \\ 
    \cline{3-4}
    & & $ResNet18$ & $82.91\%$ \\ 
    \cline{3-4}
    & & $Net$ & $83.08\%$ \\ 
    \cline{2-4}
    & \multirow{4}{*}{Ensemble-FedAvg} & $AlexNet + Net + ResNet18$ & $86.32\%$ \\
    \cline{3-4}
    & & $AlexNet^*$ & $84.67\%$ \\ 
    \cline{3-4}
    & & $ResNet18^*$ & $87.15\%$ \\ 
    \cline{3-4}
    & & $Net^*$ & $86.02\%$ \\ 
    \hline
    \end{tabular}
}
\begin{tablenotes}[para,flushleft]
* All participants use the same machine learning model structure
\end{tablenotes}
\end{table}

Table \ref{table: modelAcc} shows that the centralized setup produces slightly higher accuracy than most of the other setups when tested against the two datasets on AlexNet, ResNet18, and Net models. This is because each model receives fewer data with random distribution in the training process, making each model less generalizable compared to a centralized setup. Meanwhile, the model accuracies of our proposed method surpass the centralized setups when different models are used. This is because our proposed method utilizes ensemble weight to improve generalization and allow models with better performance to contribute more to the final results. Thus, this confirms that our proposed method does not sacrifice the accuracy of the data prediction. In fact, it increases the model accuracy compared to the existing studies. However, when a similar model structure is utilized, our proposed method produces lower accuracies compared to a centralized setup since different models can better capture a particular feature of the data compared to others. Hence, combining different models in the ensemble setup result in better accuracy.

\subsection{Ensemble-Federated Learning Model Performance}

\begin{figure}[tbh!]
    \centering
    \begin{subfigure}[tbh!]{0.2\textwidth}
        \resizebox{1\columnwidth}{!}{
            \begin{tikzpicture}
            \begin{axis}[
                xlabel={Number of images},
                ylabel={Runtime (seconds)},
                symbolic x coords= {200, 400, 600, 800, 1000},
                xtick=data,
                tick label style ={/pgf/number format/fixed},
                legend pos=south east,
                xticklabel style={anchor= east,rotate=45 },
                label style={font=\Large},
                tick label style={font=\Large},
            ]
            \addplot+[
            color=blue,
            mark size=3pt
            ]
                coordinates {
                    (200, 1.16)
                    (400, 1.23)
                    (600, 1.21)
                    (800, 1.25)
                    (1000, 1.28)
                }; 
            \end{axis}
            \end{tikzpicture}
        }
        \caption{Colon Pathology dataset}
        \label{fig: runtimeEnsColon}
    \end{subfigure}
    ~
    \begin{subfigure}[tbh!]{0.2\textwidth}
        \resizebox{1\columnwidth}{!}{
            \begin{tikzpicture}
            \begin{axis}[
                xlabel={Number of images},
                ylabel={Runtime (seconds)},
                symbolic x coords= {200, 400, 600, 800, 1000},
                xtick=data,
                tick label style ={/pgf/number format/fixed},
                legend pos=south east,
                xticklabel style={anchor= east,rotate=45 },
                label style={font=\Large},
                tick label style={font=\Large},
            ]
            \addplot+[
            color=blue,
            mark size=3pt
            ]
                coordinates {
                    (200, 0.77)
                    (400, 0.79)
                    (600, 0.81)
                    (800, 0.85)
                    (1000, 0.85)
                }; 
            \end{axis}
            \end{tikzpicture}
        }
        \caption{Breast Cancer dataset}
        \label{fig: runtimeEnsBreast}
    \end{subfigure}
    \caption{Runtime taken to evaluate and fine-tune ensemble model weights based on the number of images.}
    \label{fig: runtimeEns}
\end{figure}
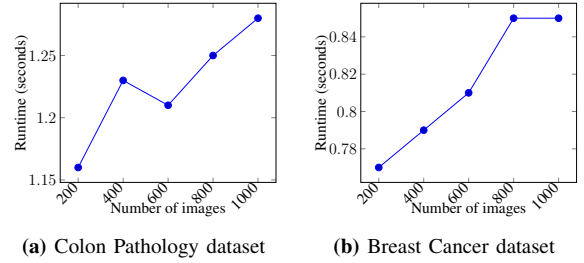

In terms of time consumption on our proposed system, Figure \ref{fig: runtimeEns} visualizes the time difference in ensemble model weights fine-tuning in regards to the datasets being used as well as the number of images used in the evaluation. Figure \ref{fig: runtimeEnsColon} and \ref{fig: runtimeEnsBreast} both indicate a negligible increase in runtime as the number of data being evaluated increases. Comparing both figures, it is clear that the time taken to determine the ensemble weights on the Colon Pathology dataset is higher than the Breast Cancer dataset. This is due to the difference in the number of classes in the classification tasks, with the Colon Pathology classification task higher than the Breast Cancer classification task.

We then evaluate the time required by different models to produce the image classification probability for ensemble model weights tuning using the SMPC protocol. Here, we also consider using a different number of images for the experiments.

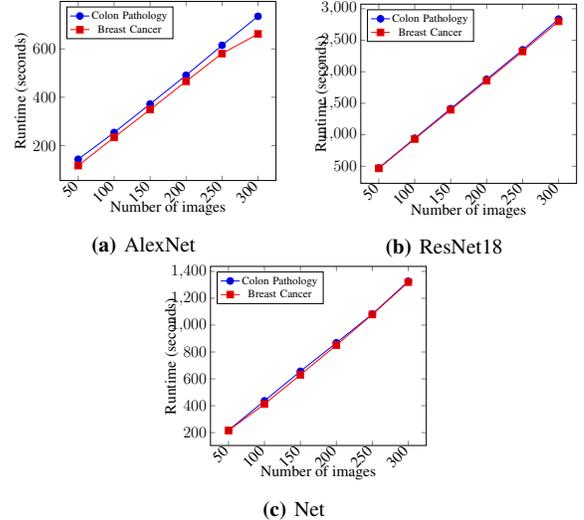
\begin{figure}[tbh!]
    \centering
    \begin{subfigure}[tbh!]{0.2\textwidth}
        \resizebox{1\columnwidth}{!}{
            \begin{tikzpicture}
            \begin{axis}[
                xlabel={Number of images},
                ylabel={Runtime (seconds)},
                symbolic x coords= {50, 100, 150, 200, 250, 300},
                xtick=data,
                tick label style ={/pgf/number format/fixed},
                legend pos=north west,
                xticklabel style={anchor= east,rotate=45 },
                label style={font=\Large},
                tick label style={font=\Large},
            ]
            \addplot+[
            mark size=3pt
            ]
                coordinates {
                    (50, 143)
                    (100, 254)
                    (150, 372)
                    (200, 491)
                    (250, 615)
                    (300, 735)
                }; 
            \addplot+[
            mark size=3pt
            ]
                coordinates {
                    (50, 117)
                    (100, 234)
                    (150, 349)
                    (200, 466)
                    (250, 581)
                    (300, 662)
                }; 
            \legend{Colon Pathology, Breast Cancer}
            \end{axis}
            \end{tikzpicture}
        }
        \caption{AlexNet}
        \label{fig: runEncAlex}
    \end{subfigure}
    ~
    \begin{subfigure}[tbh!]{0.2\textwidth}
        \resizebox{1\columnwidth}{!}{
            \begin{tikzpicture}
            \begin{axis}[
                xlabel={Number of images},
                ylabel={Runtime (seconds)},
                symbolic x coords= {50, 100, 150, 200, 250, 300},
                xtick=data,
                tick label style ={/pgf/number format/fixed},
                legend pos=north west,
                xticklabel style={anchor= east,rotate=45 },
                label style={font=\Large},
                tick label style={font=\Large},
            ]
            \addplot+[
            mark size=3pt
            ]
                coordinates {
                    (50, 473)
                    (100, 945)
                    (150, 1414)
                    (200, 1880)
                    (250, 2348)
                    (300, 2836)
                }; 
            \addplot+[
            mark size=3pt
            ]
                coordinates {
                    (50, 467)
                    (100, 933)
                    (150, 1396)
                    (200, 1858)
                    (250, 2323)
                    (300, 2800)
                }; 
            \legend{Colon Pathology, Breast Cancer}
            \end{axis}
            \end{tikzpicture}
        }
        \caption{ResNet18}
        \label{fig: runEncResnet}
    \end{subfigure}
    ~
    \begin{subfigure}[tbh!]{0.2\textwidth}
        \resizebox{1\columnwidth}{!}{
            \begin{tikzpicture}
            \begin{axis}[
                xlabel={Number of images},
                ylabel={Runtime (seconds)},
                symbolic x coords= {50, 100, 150, 200, 250, 300},
                xtick=data,
                tick label style ={/pgf/number format/fixed},
                legend pos=north west,
                xticklabel style={anchor= east,rotate=45 },
                label style={font=\Large},
                tick label style={font=\Large},
            ]
            \addplot+[
            mark size=3pt
            ]
                coordinates {
                    (50, 217)
                    (100, 436)
                    (150, 656)
                    (200, 868)
                    (250, 1082)
                    (300, 1326)
                }; 
            \addplot+[
            mark size=3pt
            ]
                coordinates {
                    (50, 215)
                    (100, 412)
                    (150, 630)
                    (200, 851)
                    (250, 1079)
                    (300, 1320)
                }; 
            \legend{Colon Pathology, Breast Cancer}
            \end{axis}
            \end{tikzpicture}
        }
        \caption{Net}
        \label{fig: runEncNet}
    \end{subfigure}
    \caption{Runtime required to produce classification probability for Colon Pathology and Breast Cancer on different CNN models using SMPC}
    \label{fig: runtimeClass}
\end{figure}

\begin{table}[!th]
\renewcommand{\arraystretch}{1.3}
\caption{Model floating-point operations (FLOPs)}
\label{table: flops}
\centering
\scalebox{0.8}{
\begin{tabular}{|c|c|c|}
\hline
Model & Dataset & G-FLOPs\\
\hline
    \multirow{2}{*}{$ResNet18$} & Colon Pathology & $113$ \\
    & Breast Cancer & $113$\\
\hline
    \multirow{2}{*}{$Net$} & Colon Pathology & $87$ \\
    & Breast Cancer & $87$\\
\hline
    \multirow{2}{*}{$Alexnet$} & Colon Pathology & $44$ \\
    & Breast Cancer & $44$\\
\hline
\end{tabular}
}% end scalebox
\end{table}

As can be seen from Figure \ref{fig: runtimeClass} runtime required during the encrypted inference to produce classification probabilities of images is more significant compared to ensemble model weights tuning as the computation cost required is higher. We further investigate the time required for different models to execute the encrypted inference process. For this, we also measure each model's computational cost by calculating the total number of floating-point operations (FLOPs) required in a single forward pass using the Keras-flops library \cite{tokusumi}. Results are shown in Table \ref{table: flops}. Further investigation on the execution time for different models revealed that models with higher FLOPs, such as the ResNet18 model, require more time than the others. While models with smaller FLOPs, such as Alexnet, require less time to complete the encrypted inference process. Hence, models with less computational cost are more suitable for participants with smaller computing power. There is no significant difference in the execution time when experiments are run on different datasets, as the number of FLOPs is not affected by the change in the datasets.

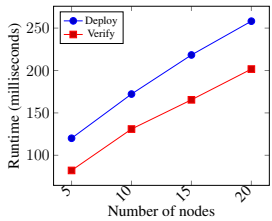
\begin{figure}[tbh!]
    \centering
    \begin{subfigure}[tbh!]{0.2\textwidth}
        \resizebox{1\columnwidth}{!}{
            \begin{tikzpicture}
            \begin{axis}[
                xlabel={Number of nodes},
                ylabel={Runtime (milliseconds)},
                symbolic x coords= {5, 10, 15, 20},
                xtick=data,
                tick label style ={/pgf/number format/fixed},
                legend pos=north west,
                xticklabel style={anchor= east,rotate=45 },
                label style={font=\Large},
                tick label style={font=\Large},
            ]
            \addplot+[
            mark size=3pt
            ]
                coordinates {
                    (5, 120.096)
                    (10, 172.214)
                    (15, 218.324)
                    (20, 258.176)			
                }; 
            \addplot+[
            mark size=3pt
            ]
                coordinates {
                    (5, 82.135)			
                    (10, 130.952)
                    (15, 165.443)
                    (20, 201.677)
                }; 
            \legend{Deploy, Verify}
            \end{axis}
            \end{tikzpicture}
        }
    \end{subfigure}
    \caption{Runtime required to perform model deployment and verification}
    \label{fig: runtimeDepVer}
\end{figure}

Finally, we provide the execution time required for the blockchain to perform data deployment to the blockchain network and data verification using the consensus mechanism. In this experiment, we used the ResNet18 model in the verification and deployment process. In Figure \ref{fig: runtimeDepVer}, we can see that as the number of blockchain nodes increases, the execution time required increases.

\begin{figure}[tbh!]
    \centering
    \begin{subfigure}[tbh!]{0.2\textwidth}
        \resizebox{1\columnwidth}{!}{
            \begin{tikzpicture}
            \begin{axis}[
                xlabel={Number of nodes},
                ylabel={Runtime (milliseconds)},
                symbolic x coords= {5, 10, 15, 20},
                xtick=data,
                tick label style ={/pgf/number format/fixed},
                legend pos=north west,
                xticklabel style={anchor= east,rotate=45 },
                label style={font=\Large},
                tick label style={font=\Large},
            ]
            \addplot+[
            mark size=3pt
            ]
                coordinates {
                    (5, 100.796)
                    (10, 120.883)
                    (15, 143.281)
                    (20, 156.552)			
                }; 
            \addplot+[
            mark size=3pt
            ]
                coordinates {
                    (5, 79.23)			
                    (10, 96.732)
                    (15, 117.224)
                    (20, 137.562)
                }; 
            \legend{Deploy, Verify}
            \end{axis}
            \end{tikzpicture}
        }
    \end{subfigure}
    \caption{Runtime required to perform ensemble weights and model id deployment and verification}
    \label{fig: runtimeDepVerMult}
\end{figure}
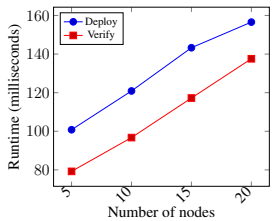

A similar trend is also shown in Figure \ref{fig: runtimeDepVerMult}, which depicts the time required for blockchain to verify and deploy hospitals' ensemble model weights and model ID. This is because data needs to be deployed to all of the blockchain nodes to be verified. Hence, the increase in the number of nodes means more data to be deployed. Each node also has to communicate with all of the blockchain nodes to ensure the hash values are identical before putting the transaction into the blockchain. Hence, more communication costs are required when the number of blockchain nodes increases, resulting in a longer execution time. From both experiments, we can see that the performance of the proposed model and ensemble model weights verification and deployment are near real-time.

\section{Conclusion}\label{sec:con}

We proposed an architecture that enables healthcare institutions to collaboratively participate in enhancing the performance of the global model while also allowing them to define their model structure. In the proposed architecture, the blockchain provides auditability and versioning control capability for the healthcare institution while also providing data integrity during local model training and ensemble model weights tuning. Our proposed model has been tested against the existing FL approach and TEE-based secure aggregation FL with blockchain. Results suggest that the proposed method is able to perform better than the existing study and able to increase the model performance effectively. We also provided empirical data on the time consumption in executing the system, namely the time required for the ensemble weights tuning, encrypted inference for model evaluation, and blockchain smart contract deployment and verification. Results show that a negligible amount of time is consumed during the ensemble weights tuning and blockchain smart contract execution. Efficiency tradeoff can be seen during the encrypted inference for model evaluation. However, higher efficiency may be achieved by FL participants with less computing power by utilizing a machine learning model with small demand for computation. This paper assumes that all hospital participants perform homogeneous tasks and train learning models with a similar model structure. The heterogeneity of the structure is considered at the hospital level. Hence, further investigation to consider heterogeneous tasks and model heterogeneity at the hospital participants' level is encouraged. Future studies should also explore methods to increase efficiency during the encrypted inference process.

\section*{Acknowledgement}
This work is supported by the Australian Research Council Discovery Project (DP210102761).
%\input{sections/06-appendix}
%\input{sections/07-acknowledgements}

% \ifCLASSOPTIONcaptionsoff
%   \newpage
% \fi
% \newpage
\bibliographystyle{IEEEtran}
\bibliography{References}

\end{document}